\documentclass[aip,apl,reprint,hidelinks]{revtex4-1}

\draft 
\usepackage[english]{babel}
\usepackage[utf8]{inputenc}
\usepackage[T1]{fontenc}
\usepackage{mathptmx}
\usepackage{graphicx}
\usepackage{dcolumn}
\usepackage{bm}
\usepackage{upgreek}
\usepackage{textcomp}
\usepackage{hyperref}
\usepackage{xcolor}

\usepackage{placeins}
\begin{document}


\title{Exfoliated hexagonal BN as gate dielectric for InSb nanowire quantum dots with improved gate hysteresis and charge noise} 



\author{Felix Jekat}
\author{Benjamin Pestka}
\affiliation{II. Institute of Physics B, RWTH Aachen University and JARA-FIT, 52074 Aachen, Germany}
\author{Diana Car}
\author{Sa\v{s}a Gazibegovi\'{c}}
\affiliation{Department of Applied Physics, Eindhoven University of Technology, 5600 MB Eindhoven, The Netherlands}
\author{Kilian Fl\"{o}hr}
\affiliation{II. Institute of Physics B, RWTH Aachen University and JARA-FIT, 52074 Aachen, Germany}
\author{Sebastian Heedt}
\altaffiliation{Current address: Microsoft Quantum Lab Delft, 2600 GA Delft, The Netherlands}
\affiliation{Peter Gr\"{u}nberg Institut (PGI-9) and JARA-FIT, Forschungszentrum J\"{u}lich, 52425 J\"{u}lich, Germany}
\author{J\"{u}rgen Schubert}
\affiliation{Peter Gr\"{u}nberg Institut (PGI-9) and JARA-FIT, Forschungszentrum J\"{u}lich, 52425 J\"{u}lich, Germany}
\author{Marcus Liebmann}
\affiliation{II. Institute of Physics B, RWTH Aachen University and JARA-FIT, 52074 Aachen, Germany}
\author{Erik P. A. M. Bakkers}
\affiliation{Department of Applied Physics, Eindhoven University of Technology, 5600 MB Eindhoven, The Netherlands}
\author{Thomas Sch\"{a}pers}
\affiliation{Peter Gr\"{u}nberg Institut (PGI-9) and JARA-FIT, Forschungszentrum J\"{u}lich, 52425 J\"{u}lich, Germany}
\author{Markus Morgenstern}
\email[]{mmorgens@physik.rwth-aachen.de}
\affiliation{II. Institute of Physics B, RWTH Aachen University and JARA-FIT, 52074 Aachen, Germany}



\begin{abstract}
We characterize InSb quantum dots induced by bottom finger gates within a nanowire that is  grown via the vapor-liquid-solid process. The gates are separated from the nanowire by an exfoliated 35\,nm thin hexagonal BN flake. We probe the Coulomb diamonds of the gate induced quantum dot exhibiting charging energies of $\sim 2.5$\,meV and orbital excitation energies up to $0.3$\,meV. The gate hysteresis for sweeps covering 5 Coulomb diamonds reveals an energy hysteresis of only $60$\,\textmu eV between upwards and downwards sweeps. Charge noise is studied via long-term measurements at the slope of a Coulomb peak revealing  potential fluctuations of $\sim$1\,\textmu eV/$\mathrm{\sqrt{Hz}}$ at 1\,Hz. This makes h-BN the dielectric with the currently lowest gate hysteresis and lowest low-frequency potential fluctuations reported for low-gap III-V nanowires. The extracted values are similar to state-of-the art quantum dots within Si/SiGe and Si/SiO${_2}$ systems.  
\end{abstract}

\pacs{}
\maketitle
Recently, nanowires (NW) of InSb and InAs \cite{Hiruma1995,Jensen2004,Bjork2002,MattiasBorg2013} came back into focus due to their large spin-orbit coupling \cite{Nilsson2009,NadjPerge2010,Fasth2007} that in combination with magnetic fields and a relatively strong proximity-induced superconductivity\cite{Doh2005,Takayanagi1985,Nilsson2012} enables tuning of Majorana modes \cite{Alicea2012,Mourik2012,Albrecht2016,Lutchyn2018} as a basis for topologically protected quantum computing. \cite{Stern2013,Vijay2015,Litinski2017} Typically, the NWs are tuned electrically by a number of bottom finger gates that are separated from the NW by a gate dielectric.\cite{Mourik2012,Das2012} It is well known that both charge noise and hysteresis of gate-induced potentials deteriorate the performance of semiconductor qubits, \cite{Culcer2009,Kuhlmann2013,Yoneda2017,Mi2018} as is also expected for the prospective Majorana qubits.\cite{Schmidt2012,Li2018} Hence, it is crucial to optimize the dielectric in terms of unintentional charge fluctuations. 

For exfoliated two-dimensional materials such as graphene, it turned out that hexagonal boron nitride (h-BN) is ideal for that purpose. \cite{Geim2013,Frisenda2018} For example, it improves the charge carrier mobility by more than an order of magnitude compared to the previously used Si/SiO$_2$.\cite{Dean2010,Banszerus2016} Furthermore, it is easy to fabricate. Thus, exploiting exfoliated h-BN as gate dielectric for low-gap III-V NWs is  appealing.  First experiments used h-BN to separate the global Si/SiO$_2$ back gate from an InSb NW enabling the first quantized conductance steps in such NWs at zero magnetic field.\cite{Kammhuber2016} Subsequently, measurements on  proximity-coupled InSb NWs on h-BN showed magnetic field induced zero bias peaks, indicative of the presence of Majorana zero modes.\cite{Guel2018, Gill2018}  However, Coulomb diamonds (CDs) with excited states in a gate-induced quantum dot (QD) have not been reported and, more importantly, the charge noise and gate hysteresis of such NWs on  h-BN have not been studied. Reports on these properties are only available for other types of dielectrics. \cite{Sakr2008,Nilsson2008, Delker2012,Vitusevich2017,Delker2013,Petrychuk2019,Persson2013,Persson2010,Wahl2013,Volk2010,Guel2015} They exhibit, e.g., a relatively large  low-temperature gate hysteresis on LaLuO$_3$ and SiO$_2$ being  $0.5$\,V and $ 2$\,V at gate sweep ranges of 4\,V and 30\,V, respectively.\cite{Volk2010,Guel2015} Noise properties for QDs have only been reported for a vacuum dielectric revealing $1/f$ behavior above $\sim$ 300\,Hz and an upturn at lower frequency with noise of $\sim 0.2$\,µeV$/\mathrm{\sqrt{Hz}}$ at 100 Hz.\cite{Nilsson2008}

 Here, we study an InSb NW/h-BN device with bottom finger gates (pitch 90\,nm) at the temperature  $T=300$\,mK. The device exhibits a gate hysteresis of 2\,mV for sweeps of 150\,mV (250\,mV) at a rate of 25\,mV/s (42\,mV/h), hence, significantly better than in previous reports.\cite{Volk2010,Guel2015}  It, moreover, shows a charge noise of only 1\,µeV$/\mathrm{\sqrt{Hz}}$ at $ 1$\,Hz with an approximate $1/f^{1.5}$ dependence towards lower frequencies. The noise is similar to the previously studied vacuum dielectric\cite{Nilsson2008} pointing to  remaining limitations due to defects at the NW itself. More importantly, the value is slightly better than for state-of-the-art QDs in Si/SiGe or Si/SiO${_2}$ structures ($\sim 3$\,µeV$/\mathrm{\sqrt{Hz}}$ at $1$\,Hz). \cite{Freeman2016,Petit2018,Connors2019,Mi2018} 
Hence, h-BN turns out to be a favorable dielectric for low-gap III-V NWs.

\begin{figure*}
\includegraphics[width=0.99\textwidth]{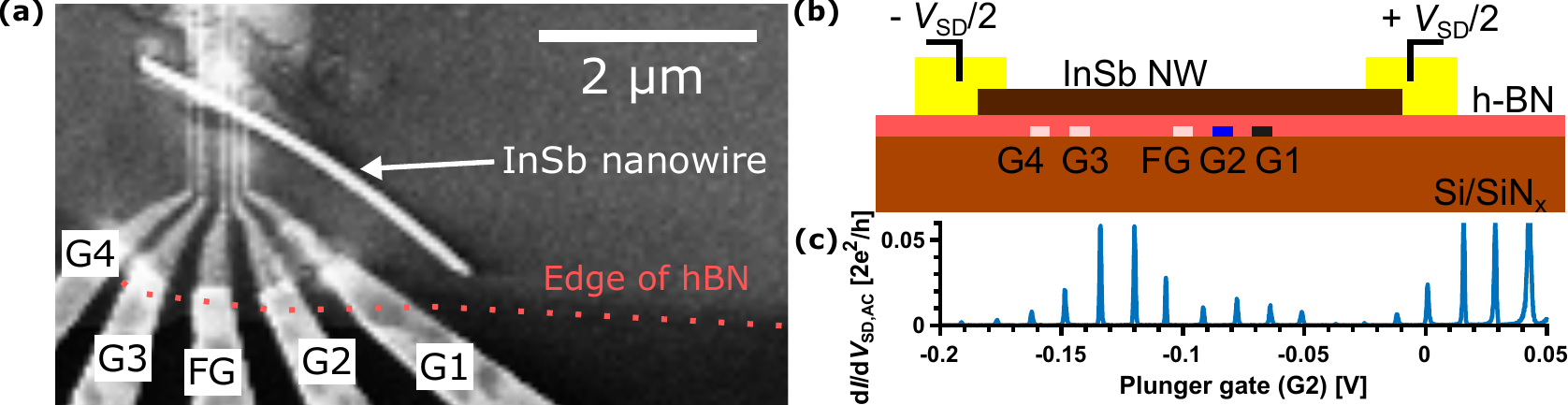}
\caption{\label{fig:fig1} (a) Scanning electron microscopy image of the device prior to depositing source and drain contacts. Different finger gates (G1-G4) and a floating gate (FG) are labeled. NW diameter: $\sim 100$\,nm. (b) Side view sketch of the device. Si/SiN$\mathrm{{_x}}$ (light brown) acts as a global backgate (BG). Finger gates G1-G4 and FG (white, blue, black) are deposited on top and buried below a 35\,nm thick h-BN flake (pink). The NW (dark brown) is contacted by two Ti/Au pads (yellow) at distance 2.2\,\textmu m used to apply the source-drain voltage $V_{\rm SD}$. Light brown, blue and black colors of gates match the colors of the corresponding phase stability diagrams in Fig.~\ref{fig:fig2}, where these gates are used as plunger gate.   (c) NW conductance d$I$/d$V_{\rm SD,AC}$ as a function of plunger gate voltage $V_{\rm G2}$ at gate G2. Coulomb peaks appear due to the formation of a QD confined by energy barriers that are induced via G1 and G3, $V_{\rm G1}=-970$\,mV,  $V_{\rm G3}=-463$\,mV, $V_{\rm BG}=3$\,V, $V_{\rm SD,AC}=20$\,\textmu V, $f_{\rm AC}= 933.5$\,Hz, $V_{\rm SD,DC}=0$\,V,   $T=300$\,mK.} %
\end{figure*}

The InSb NWs were grown on top of InP stems via the vapor-liquid-solid (VLS) method using a gold droplet as catalyst.\cite{Plissard2012,Car2014} A QD device of such a NW (Fig.~\ref{fig:fig1}(a)$-$(b)) consists of a 200\,nm thick SiN$\mathrm{{_x}}$ layer, on a highly doped Si substrate acting as a global back gate (BG) with multiple finger gates (G1$-$G4, FG) on top. The finger gates are 35\,nm wide and defined by electron beam lithography (EBL) with a spacing of 55\,nm except between G3 and FG where the spacing is 130\,nm. An h-BN flake is deposited on top of the finger gates via the dry transfer method.\cite{Dean2010} Subsequently, one InSb NW is placed onto the h-BN with sub-\textmu m lateral precision via an indium tip attached to a micromanipulator. \cite{Floehr2011} Finally,  source and drain contacts are prepared via EBL. Prior to the metal deposition of the Ti/Au (10\,nm/110\,nm) contacts, the exposed NW area is passivated ex-situ by sulphur\cite{Suyatin2007} and subsequently cleaned in-situ by argon ion bombardment. Transport measurements are performed at  $T=300$\,mK in a $^3$He magneto-cryostat (Teslatron from Oxford Instruments).  Before cool-down, the insert is evacuated to $10^{-6}$\,mbar for 48\,h at 300\,K in order to remove adsorbates from the NW surface.\cite{Guel2015} 
\begin{figure*}
\includegraphics[width=0.99\textwidth]{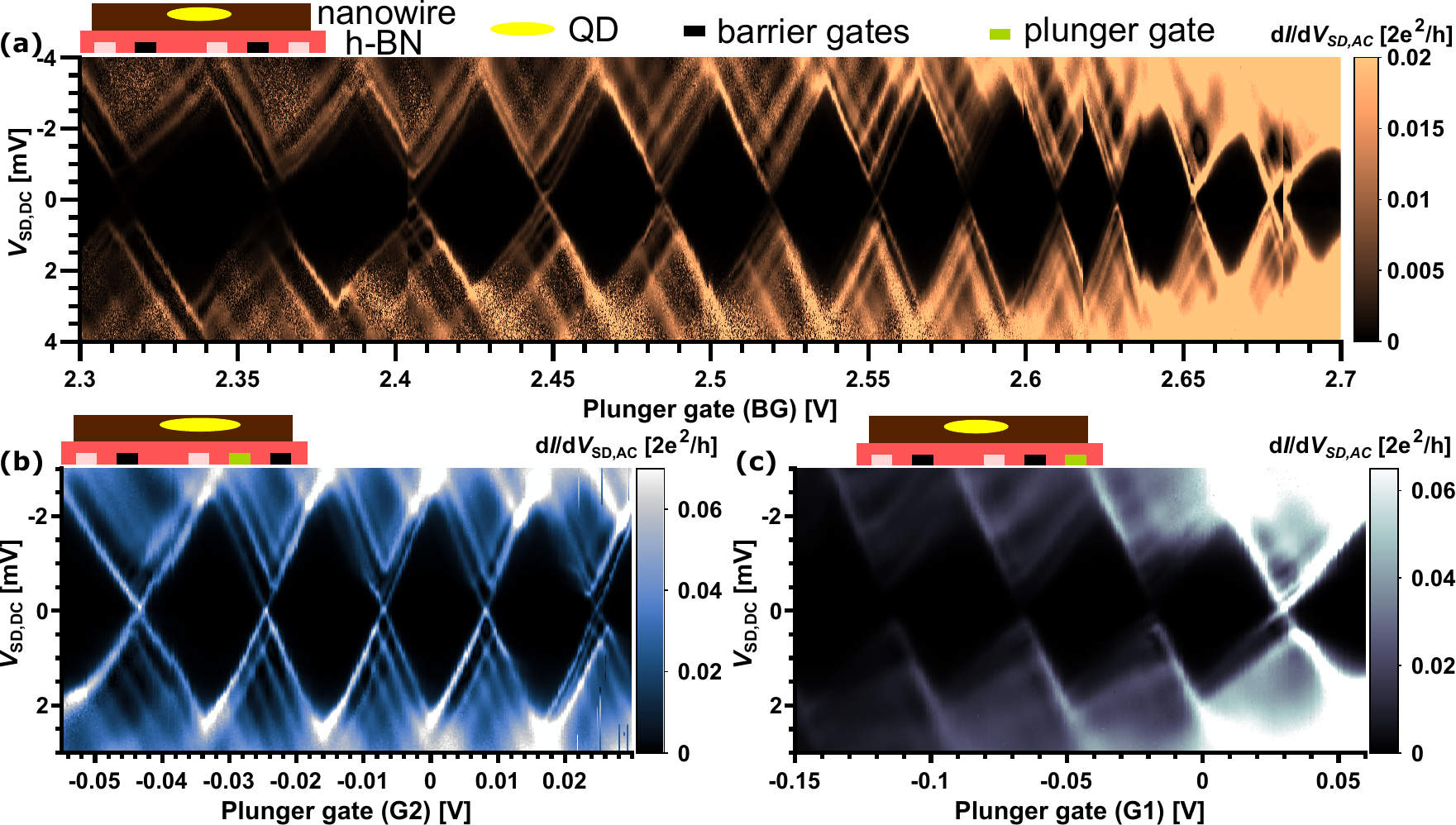}

\caption{\label{fig:fig2} Charge stability diagrams of different QDs in the same InSb NW using different combinations of finger gates as sketched on top including the resulting QD area (yellow). Fast scan direction is along $V_{\rm SD, DC}$, $V_{\rm SD,AC}=20$\,\textmu V, $f_{\rm AC}= 83$\,Hz, $V_{\rm G4}=0$\,V, $T=300$\,mK. (a) QD confined by G2 and G3 (black) and charged by the back gate (BG), $V_{\rm G2} =-700$\,mV, $V_{\rm G3}=-1$\,V, $V_{\rm G1}=0$\,V.  A few perturbations are visible at $V_{\rm BG}=2.68$\,V, 2.64\,V, 2.62\,V and 2.6\,V, likely due to uncontrolled charging events during the early stages of the measurement.  (b) QD confined by G1 and G3 (black) and charged by G2 (green), $V_{\rm G1} =-650$\,mV, $V_{\rm G3}=-980$\,mV, $V_{\rm BG}=3$\,V. (c) QD confined by G2 and G3 (black) and charged by G1 (green), $V_{\rm G2} =-580$\,mV, $V_{\rm G3}=-922$\,mV, $V_{\rm BG}=3$\,V.}%
\end{figure*}

Gate dependent conductivity traces (not shown) reveal a low temperature mobility of the NW of $\mu=28000$\,cm$^2$/Vs.\cite{Guel2015} 
Using the finger gates, we induce a QD within the NW exhibiting regularly spaced Coulomb peaks of different heights (Fig.~\ref{fig:fig1}(c)), probably due to different coupling of the states to the tunnel barriers.   
Different combinations of finger gates reveal charge stability diagrams of such QDs with regularly spaced CDs for all combinations of gates and excited states at larger $V_{\rm SD}$ (Fig.~\ref{fig:fig2}(a)-(c)). Only very few perturbations appear, likely caused by uncontrolled charging events in the surrounding of the QD. We could not measure the last CD prior to depletion probably due to the elongated QD geometry decoupling the lowest energy state from the tunnel barriers.   
For the CDs of Fig.~\ref{fig:fig2}(a)-(c), one straightforwardly deduces charging energies $E_{\rm C}$ up to 3\,meV, 2.3\,meV, and 2.5\,meV and lever arms $\alpha$
of 0.05\,eV/V, 0.12\,eV/V, and 0.03\,eV/V, respectively. Estimating the QD extension via the QD capacitance  $C=e^2/E_{\rm C}=70$\,aF assuming, for the sake of simplicity, charging of an isolated sphere of radius $r$, we reasonably find $2r=2\frac{C}{4\pi\epsilon_0\epsilon_r}=74$\,nm \cite{Shorubalko2007} using the dielectric constant of InSb $\epsilon_r=16.8$ . The deduced diameter $2r$ is a bit smaller than gate spacing and NW diameter ($\sim 100$\,nm). This could partly be due to squeezing of the QD area in the direction perpendicular to the wire axis via the gate voltages as indeed found by COMSOL\textsuperscript{\textcopyright} simulations.\cite{Heedt2015}

%
\begin{figure*}
\includegraphics[width=0.99\textwidth]{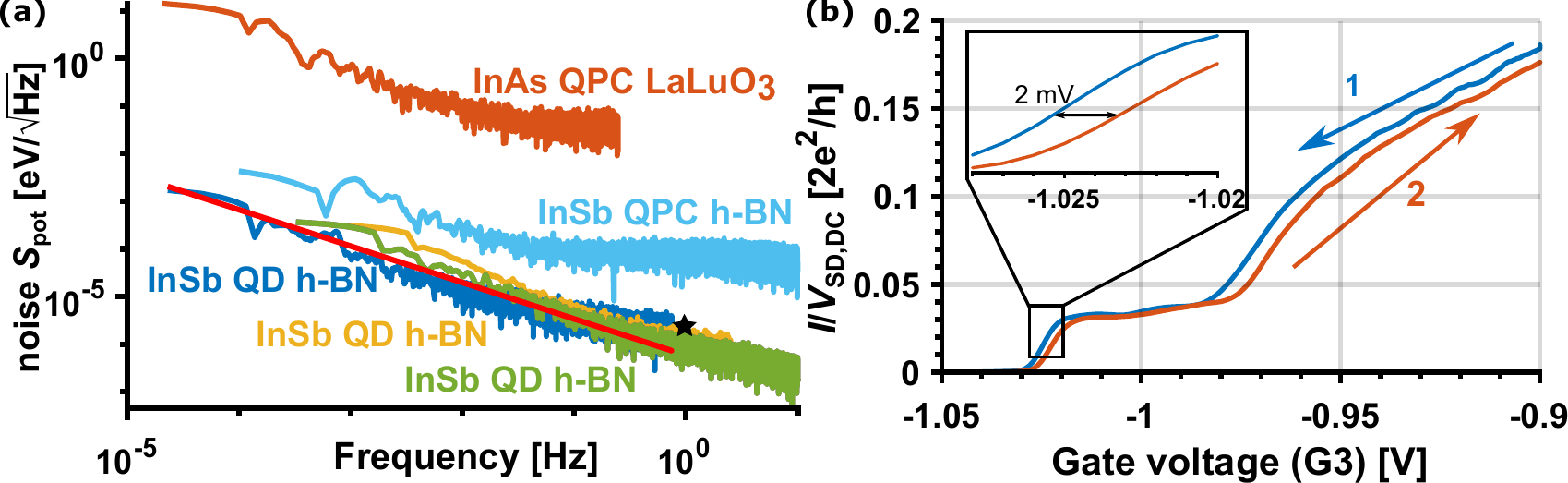}

\caption{\label{fig:fig3}(a) Potential fluctuation noise $S_{\rm pot}(f)$ as function of frequency $f$ for several InSb NW QDs on h-BN using different gate configurations that yield $E_{\rm C}=3$\,meV (blue), $E_{\rm C}=1.8$\,meV (yellow), $E_{\rm C}=1$\,meV (green) with fit curve $S_{\rm pot}^2(f) \propto 1/f^{1.5}$ (red). The black star shows a benchmark $S_{\rm pot}(f)$ for Si and Si/SiGe QDs\cite{Petit2018,Connors2019,Freeman2016}.
Additionally, $S_{\rm pot}(f)$ for a QPC within the same InSb NW on h-BN (light blue, $V_{\rm SD,DC}=3$\,mV, $V_{\rm SD,AC}=20$\,\textmu V, $f_{\rm AC}= 933.5$\,Hz, $V_{\rm G3} =-600$\,mV, all other gates grounded)  is compared to $S_{\rm pot}(f)$ of a QPC within an InAs NW on LaLuO$_3$ (orange, $V_{\rm SD,DC}=5$\,mV, $V_{\rm SD,AC}=20$\,\textmu V, $f_{\rm AC}= 1.1$\,kHz, $V_{\rm G1} =-10$\,V, all other gates grounded). Both devices are fabricated using the same deposition methods except for the dielectric. The noise background as determined in the Coulomb blockade region of the QDs is more than a factor of 10 lower than all displayed $S_{\rm pot}(f)$. (b) $V_{\rm G3}$ dependent conductance of the InSb NW on h-BN close to pinch-off (without inducing a QD). Arrows with numbers indicate the subsequent sweep directions of $V_{\rm G3}$. Inset: zoom showing a hysteresis of $\sim 2$\,mV, sweep rate: 25\,mV/s, $V_{\rm SD,DC}=3$\,mV, $V_{\rm SD,AC}=20$\,\textmu V, $f_{\rm AC}= 83$\,Hz, $V_{\rm BG}=3$\,V, all other gates grounded. (a), (b)  $T=300$\,mK.}%
\end{figure*}

To quantify the charge noise acting on the NW QDs, Fig.~\ref{fig:fig3}(a) shows low-frequency noise measurements for the three different gate configurations. We measure the temporal current fluctuations $\delta I(t)$ at the slope of a Coulomb peak for $V_{\rm SD,AC}=20$\,\textmu V. In order to transfer this to the potential fluctuation noise $S_{\rm pot}(f)$ as function of frequency $f$, we firstly use the measured shape of the Coulomb peak in $I(V_{\rm Gate})$ traces, well fitted by a Fermi-Dirac peak, to deduce the gate voltage variation, $\delta V_\mathrm{Gate}(t)$. Then, we transfer $\delta V_\mathrm{Gate}(t)$ to potential energy variation $\delta E(t) = \alpha \delta V_\mathrm{Gate}(t)$ with $\alpha$ as deduced from respective CDs. The square root of the single-sided power spectral density of the resulting  $\delta E(t)$ in the QD leads to $S_{\rm pot}(f)$ in eV/$\mathrm{\sqrt{Hz}}$ as displayed in Fig.~\ref{fig:fig3}(a) with the rms potential noise being $\delta E_\mathrm{rms}^2 = \int S_{\rm pot}^2 (f)\, df$ across the measurement bandwidth. 
We find $S_{\rm pot}(1$\,Hz$)=1$\,µeV$/\mathrm{\sqrt{Hz}}$ and an increase towards lower $f$ mostly following $S_{\rm pot}^2(f) \propto 1/f^{1.5}$ (red fit line). The enhanced logarithmic slope of $S_{\rm pot}^2(f)$ with respect to the classical $1/f$ noise is in reasonable agreement with the upturn of the $1/f$ noise below $f=100$\,Hz observed earlier for InAs NWs with vacuum dielectric.\cite{Nilsson2008} 

We also display a direct comparison of quantum point contacts (QPCs) for the InSb NW on hBN with an InAs NW\cite{Do2007} on a LaLuO$_3$ dielectric. Except for the dielectric, deposited via pulsed laser deposition \cite{Volk2010}, the two devices are prepared identically.  The QPC is formed by charging one of the finger gates only with all other gates grounded. The displayed $S_{\rm pot}(f)$ (Fig.~\ref{fig:fig3}(a), orange) originates from $\delta I(t)$ at the pinch-off of the NW induced by a single finger gate. It is converted to $S_{\rm pot}(f)$ by the measured $I(V_{\rm Gate})$ using
$\alpha_{\rm hBN}=0.12\pm0.02$\,eV/V for the hBN device as determined from the corresponding QD CDs with error bars as deduced from CD variations and $\alpha_{\rm LaLuO_3} =0.55\pm 0.11$\,eV/V for the LaLuO$_3$ device deduced by scaling $\alpha_{\rm hBN}$ for the different thicknesses (hBN: $35 \pm 2$\,nm, LaLuO$_3$: $50\pm 5$\,nm) and for the different $\epsilon$ (hBN: $\simeq 4$ \cite{Young2012}, LaLuO$_3$: 26$\pm 1$ \cite{Schfer2015}).
Remarkably,  $S_{\rm pot}(f)$ of the h-BN device is more than two orders of magnitude lower than for the LaLuO$_3$ device (Fig.~\ref{fig:fig3}(a)) illustrating the excellent properties of the hBN dielectric.  We employed all four finger gates for such QPC measurements. The resulting $S_{\rm pot}(f)$ curves are nearly identical up to 1\,Hz, but vary at higher frequency being either lower by up to a factor of four or larger by  up to a factor of three with respect to the displayed curve. This indicates the presence of particular fluctuators at $\sim 10$\,Hz in the device \cite{chanrion2020,Connors2019}.
 
 \begin{figure*}
\includegraphics[width=0.99\textwidth]{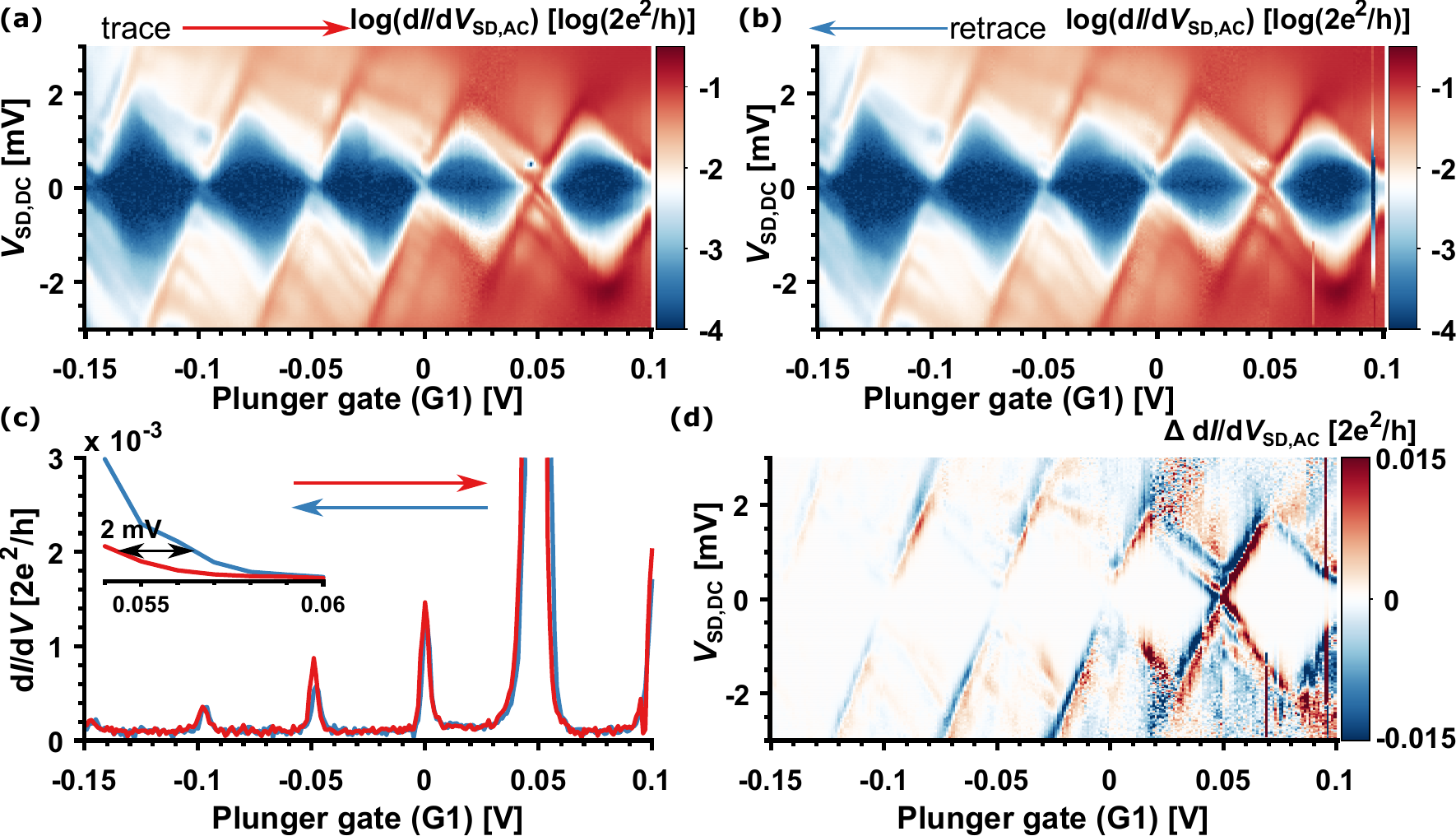}
\caption{\label{fig:fig4}(a), (b) Charge stability diagrams for a QD confined by gates G2 and G3 and charged by gate G1. The two diagrams are recorded directly after each other with different gate sweep directions as marked by arrows on top. Fast sweep direction is along $V_{\rm SD,DC}$, total measurement time: 12\,h, gate sweep rate: 42 mV/h. (c) Line cut through (a) (red) and (b) (blue) at $V_{\rm SD,DC}=0$\,mV. Inset: zoom with marked hysteresis of 2\,mV. (d) Difference of the data of (b) and (a). Parameters: $V_{\rm G2}=-580$\,mV, $V_{\rm G3}= -922$\,mV, $V_{\rm BG}=3$\,V, $V_{\rm SD,AC}=20$\,\textmu V, $f_{\rm AC}= 83$\,Hz, $T=300$\,mK.}%
\end{figure*}

 Comparison with literature data on noise for III-V low-gap NWs is difficult. Either much longer parts of the NW are gated\cite{Sakr2008}, effectively averaging charge fluctuations, or the  frequencies are larger due to probing by radio frequency via reflection at the QD. \cite{Nilsson2008} 
 Extra\-po\-lating the latter noise data obtained for a suspended InAs NW (vacuum dielectric) at 100\,Hz\cite{Nilsson2008} towards $10$\,Hz via the measured $1/f^{1.5}$ dependence leads to $S_{\rm pot}(10$ Hz$)\simeq 1$\,\textmu eV$/\mathrm{\sqrt{Hz}}$, larger than for our device on h-BN ($ 0.4 $\,\textmu eV$/\mathrm{\sqrt{Hz}}$ at 10 Hz). However, the data with vacuum dielectric are measured at 1.5\,K such that extrapolating to 0.3\,K by the established linear temperature dependence\cite{Petit2018,Connors2019} yields $S_{\rm pot}(10$ Hz$, 0.3\,K)\simeq 0.2$\,\textmu eV$/\mathrm{\sqrt{Hz}}$, slightly better but still rather similar to our device. Since vacuum exhibits no defects acting as charge traps, Nilson et al.\cite{Nilsson2008} 
conjectured that the major charge noise originates from charge traps within or on the NW and not from the dielectric. Regarding the similarity of $S_{\rm pot}(10$ Hz$, 0.3\,K)$, we believe that this is correct for our device, too.
 
  It is instructive to compare our data with the charge noise in Si or Si/SiGe QDs\cite{Connors2019,Freeman2016,Petit2018, Mi2018, chanrion2020}, currently considered as most promising for semiconductor spin qubits\cite{Watson2018}. For these QDs, one finds $S_{\rm pot}^2(f\simeq 1$\,Hz$) \propto 1/f^\beta$ with device-dependent $\beta = 1-1.4$ and, consistently, an increase of $S_{\rm pot}(f)$ with increasing $T$.  Favorably, the reported $S_{\rm pot}(f)$ at 0.3\,K (1-5\,\textmu eV$/\mathrm{\sqrt{Hz}}$ at 1\,Hz)\cite{Petit2018,Connors2019,Freeman2016,chanrion2020} (star in Fig.~\ref{fig:fig3}(a))  is not smaller than for our InSb NW QD on h-BN ($\sim 1$\,\textmu eV$/\mathrm{\sqrt{Hz}}$ at 1\,Hz).  
 This renders the device competitive to the most favorable material combinations in terms of charge noise.

The second important benchmark for a dielectric is the gate hysteresis. Figure~\ref{fig:fig3}(b) reveals that the InSb NW on h-BN exhibits a gate hysteresis $\Delta V_{\rm hyst}\simeq 2$\,mV (inset) at a sweep rate of 25\,mV/s. Since it is known that the hysteresis strongly depends on the probed gate range, we scale the hysteresis by the gate range for comparison.
For the gate range $\Delta V_{\rm gate}=150$\,mV, this leads to a ratio $R=\Delta V_{\rm hyst}/\Delta V_{\rm gate}= 0.013$. This is much lower than observed previously for InAs or InSb NWs on other gate dielectrics: $R=0.13$\cite{Volk2010}, $R=0.07$\cite{Guel2015}, where
higher temperatures (25\,K, 4.2\,K) have been employed that typically even reduce hysteresis as indeed found in one of these studies \cite{Volk2010}.
Since $R$ additionally depends on the sweep rate\cite{Lynall2018,Dayeh2007}, we improved it further by reducing the gate sweep rate to 42\,mV/h leading to $R=0.008$ with $\Delta V_{\rm gate}=250$\,mV. The extremely low rate is employed to record full charge stability diagrams subsequently for both gate sweep directions (Fig.~\ref{fig:fig4}(a),(b)). The  total measurement time of 12\,h evidences the long term stability of the QD by the excellent similarity of the two diagrams.  Only two conductivity jumps  (Fig.~\ref{fig:fig4}(b), right) are observed. The gate hysteresis is quantified by a line cut at $V_{\rm SD}=0$\,mV (Fig.~\ref{fig:fig4}(c)) revealing  $\Delta V_{\rm hyst}=2$\,mV as maximum hysteresis between the two curves (inset) and, hence, implying $R=0.008$. Using $\alpha=0.03$\,eV/V, one can calculate the energy hysteresis $\Delta E=60$\,\textmu eV. Figure~\ref{fig:fig4}(d) displays the difference between Fig.~\ref{fig:fig4}(b) and Fig.~\ref{fig:fig4}(a) showing that the small hysteresis is reliably observed across the whole charge stability diagram. The observation of a small hysteresis, low charge noise and a small number of jumps in stability diagrams consistently indicate a very low number of chargeable impurities in the h-BN layer, thus making it a favorable dielectric for III-V NW devices. As pointed out above, the performance is likely limited by the remaining charge traps on the NW itself.

In summary, we have presented an InSb NW device with an h-BN flake as gate dielectric. With  a set of finger gates, electrons are confined in QDs using different gate configurations, resulting in regular Coulomb diamonds with multiple excited states. Favorably, the device has the lowest noise level  (1\,\textmu eV/$\mathrm{\sqrt{Hz}}$ at $\sim 1$\,Hz) reported for low-gap III-V NW devices yet and shows an unprecedented gate hysteresis of 2\,mV only. Hence, in terms of charge noise, h-BN is the currently most favorable dielectric for low-gap III-V NW devices.
\FloatBarrier


%
%
%

\begin{acknowledgments}
We thank S. Trellenkamp and F. Lentz for help with the EBL and H. Bluhm for helpful discussions. Funding by the German Research Foundation (DFG) under Germany's Excellence Strategy – Cluster of Excellence Matter and Light for Quantum Computing (ML4Q) EXC 2004/1 – 390534769 and by European Graphene Flagship Core 2, grant number 785219 is gratefully acknowledged.
\end{acknowledgments}
\section*{Data availability}
The data that support the findings of this study are available from the corresponding author upon reasonable request.
\bibliography{FelixBib}

\end{document}